\documentclass[prl,twocolumn]{revtex4}         
\usepackage{graphicx}
\newcommand{\ket}[1]{\mbox{\ensuremath{|#1\rangle}}}
\newcommand{\bra}[1]{\mbox{\ensuremath{\langle#1|}}}

\begin{document}
\title{Absolute emission rates of Spontaneous Parametric Down Conversion into
  single transverse Gaussian modes} 
\author{Alexander Ling$^{1,2}$, Ant\'{\i}a Lamas-Linares$^2$ and Christian Kurtsiefer$^2$}
\affiliation{$^1$Temasek Labs, National University of Singapore, Singapore 117508}
\affiliation{$^2$Centre for Quantum Technologies / Department of Physics,
  National University of Singapore, Singapore 117542}
\email[]{christian.kurtsiefer@gmail.com}

\homepage[]{http://www.quantumlah.org}
\date{\today}

\begin{abstract}
We provide an estimate on the absolute values of the emission rate of photon
pairs produced by spontaneous parametric down conversion in a bulk crystal
when all interacting fields are in single transverse Gaussian modes. Both
collinear and non-collinear configurations are covered, and we arrive at a
fully analytical expression for the collinear case.
Our results agree reasonably well with values found in typical experiments,
which allows this model to be used for understanding the dependency on the
relevant experimental parameters.
\end{abstract} 
\maketitle

\section{ Introduction}
For the last two decades, spontaneous parametric down-conversion (SPDC) has
been the workhorse process for the generation of correlated photon
pairs. These can easily be cast into maximally entangled states, which are
useful e.g. for violating Bell inequalities \cite{eprspdc}, or for the
investigation of other fundamental aspects of quantum mechanics. Progress in
SPDC-based photon pair sources has allowed the emitted pairs from such sources
to play a leading role in demonstration of quantum information techniques, and
made its way to almost practical applications like quantum key distribution.

A number of studies have established the basic understanding of SPDC
\cite{othertheories,theories,klyshkotxt}, based on energy conservation, and
momentum conservation when participating light fields are treated as plane
waves \cite{hong85}.

Many of the more recent applications often necessitate significant
manipulation and transport of the photon pairs; this can be achieved in a
convenient way by guiding the light in single mode optical fibers.
The basic idea of modeling SPDC in this regime is to map the optical
modes propagating in the fibers into freely propagating modes of the
electromagnetic field in the nonlinear conversion material, where they
interact with a pump field. These freely propagating spatial modes can be
described in good approximation by paraxial Gaussian beams, and any
optimization strategy will involve some sort of mode matching of such
interacting beams.

Previous studies of SPDC light coupled into single mode fibers have focused in
optimizing the coupling efficiency, defined as the ratio of photon pairs to
single photons that are observed because this is a quantity which can be
measured easily in an experiment. This quantity is important for developing
loophole-free tests of Bell's inequality \cite{monken98}, heralded single
photon sources \cite{bovino03,expcoupling,ljunggren05}, or simply sources of
high pair brightness \cite{kurtsiefer01}.

So far, theoretical work in this area has focused mostly in such secondary
parameters, and no closed expression for the absolute rate of photon pairs was
available for typical experimental configurations. This made it
difficult to estimate whether a particular experimental source implementation
could be improved with respect to a particular figure-of-merit, be it total
rate or spectral brightness.

In this paper, we try to address this problem and derive an expression for the
absolute rate of SPDC emission from a bulk crystal into Gaussian modes.
The work connects to earlier investigations of absolute SPDC rates with
beams of finite diameter by Kleinman and Klyshko \cite{theories, klyshkotxt}.
It was found there that the overall rate of pair production is independent of
the spot-size of the pump beam \cite{theories}, and that the conversion
efficiency of pump photons into correlated pairs integrated over all emission
directions \cite{klyshkotxt} is in the order of $10^{-8}$ per mm for a typical
non-linear material.  
The restriction on specific spatial modes defined by single mode optical
fibers in the more recent applications, however, made it difficult to relate
their results directly to experiments. Our description applies
both to Type I and II phase matching  conditions, and covers collinear and
non-collinear geometries important for the generation of
polarization-entangled photon pairs \cite{kwiat95,kwiat99}. 
 
\begin{figure} \centerline{\scalebox{0.28} { \includegraphics{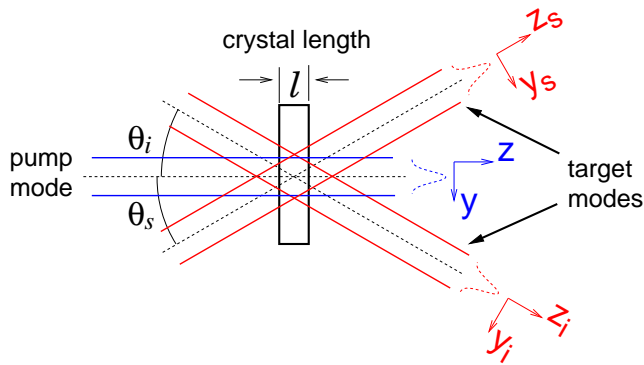}}}
  \caption{\label{fig:concept} (Color online) Schematic of the down conversion
    model 
    considered in this paper.  Pump, signal and idler beams are treated
    as paraxial beams with a Gaussian transverse mode. The x-axis is
    perpendicular to the plane of the diagram.  Coordinate systems of the
    signal and idler ($x_{s,i},y_{s,i},z_{s,i}$) are tilted by angles
    $\theta_s, \theta_i$ with respect to the coordinate system of the pump.}
\end{figure}

\section{Model}
The basic process of SPDC can be understood as the spontaneous decay of a
photon from a pump field into two daughter photons propagating in two
---possibly different--- target modes, with the process being mediated by a
material with a nonlinear optical susceptibility. The physical implementation
of SPDC utilizes the lowest order of the nonlinear susceptibility tensor
$\chi^{(2)}$ in an appropriate material.  While energy conservation allows the
decay process to take place in many target modes, phase matching requirements
need to be engineered to allow conversion to take place into any particular
pair of directions.

The physical model of the three interacting optical modes is depicted in
figure~\ref{fig:concept}. We treat the pump beam and the target modes
for the down-converted light as propagating paraxial beams with a
Gaussian transverse profile. The beams overlap within a nonlinear optical
crystal of finite length $l$, with surfaces normal to 
the propagation direction of the pump beam. Pump and target modes propagate
in one plane, but need not to be parallel. We further assume that the three
interacting modes overlap in a region without a significant variation of the
transverse profile along their respective propagation directions.
This is a reasonable assumption for typical Gaussian beam parameters and conversion crystal lengths used in experiments \cite{kurtsiefer01}. 

We follow the tradition of referring to the target modes as signal (index $s$)
and idler (index $i$) and choose coordinate systems where the $z_{s,i,p}$
directions are parallel to the main propagation direction for each mode $
s,i,p$ ($p$ refers to the pump mode). 
The spatial mode function of the electrical field for each of the modes can be
written as 
\begin{equation}
g({\bf r}) = e^{i kz}\cdot U(x,y) = e^{i k z}\cdot e^{-{ x^2+y^2\over W^2}}\quad,
\end{equation}
where $k$ denotes the $z$-component of the corresponding wave vector, 
$W$ the Gaussian beam waist parameter, and $x,y,z$ refer to the
corresponding coordinate system for each mode.
To simplify the overlap calculations, we use normalization constants
$\alpha$ for the envelope functions $U(x,y)$
such that
\begin{equation}\label{eq:normalize1}
\alpha^2\int \,dxdy \left|U(x,y)\right|^2=1
\end{equation}
in their corresponding coordinate systems, which implies
\begin{equation}
\alpha_{p,s,i}=\sqrt{2\over\pi W_{p,s,i}^2}\,.
\end{equation}

We note that the spatial mode function $g({\bf r})$ fulfills Maxwell's
equations only approximately. For the calculations presented below, however,
this poses no problem. 
Furthermore, the dispersion relation connected with this mode function has the confinement correction:
\begin{equation}
\omega^2 = c^2\left( k^2+ \frac{2}{W^2} \right)
\end{equation}
Again, for practical beam diameters $W$ of about 100 wavelengths considered
in this paper, this correction term is small enough to safely neglect it. 

\subsection{Pump mode}
The pump mode is aligned with the main coordinate system
$x,y,z$, and treated as a classical monochromatic field of amplitude
$E_p^0$. We further assume (as is customary) that we have no significant
depletion of the pump 
in the downconversion crystal. The electrical field of the pump can thus be
written as 
\begin{eqnarray}\label{eq:pumpmode}
  {\bf E}_p({\bf r},t) &=& {1\over 2}\left[ {\bf E}_p^{(+)}({\bf r},t) + {\bf E}_p^{(-)}({\bf r},t)\right] \\ \nonumber
  &=&{1\over2}\left[E_p^0 {\bf e}_p g_p({\bf r}) e^{-i\omega_pt}+c.c.\right],
\end{eqnarray}
with a polarization vector ${\bf e}_p$, and a corresponding angular
frequency $\omega_p$. Using the normalization expression
(\ref{eq:normalize1}), we can connect the electrical field amplitude $E_p^0$
with the optical power $P$ in the pump beam:
\begin{equation}
\left|E_p^0\right|^2=\alpha_p^2{2 P\over\epsilon_0 n_p c}\quad,
\end{equation}
with the refractive index $n_p$ for the pump field, the electrical field
constant $\epsilon_0$ and the speed of light $c$ in vacuum.

\subsection{Target modes}
First, we need to take care of the possibly non-collinear propagation
of the target modes with respect to the pump. By introducing target mode angles
$\theta_{s,i}$, and using an orientation as indicated in
figure~\ref{fig:concept}, we express the spatial coordinates of the target
modes in terms of the main coordinates $x,y,z$:
\begin{equation}\label{eq:coordtransf}
  \left(
    \begin{array}{c} x_{s,i} \\ y_{s,i} \\ z_{s,i} \end{array}
  \right) = 
  \left(
    \begin{array}{ccc}1 & 0 & 0 \\ 
      0 & \mbox{cos}\theta_{s,i} & \pm\mbox{sin}\theta_{s,i} \\
      0 & \mp\mbox{sin}\theta_{s,i} & \mbox{cos}\theta_{s,i} 
    \end{array} \right)
  \left(
    \begin{array}{c} x \\ y \\ z \end{array}
  \right)
\end{equation}

To arrive at a rate of photon pairs generated via SPDC, we use the
quantized field operators. We do that by introducing a 
quantization length $L$ in propagation direction for clarity in
the counting of modes, and postulate periodic boundary conditions; later we
will drop this requirement. Following the notation in equation
(\ref{eq:pumpmode}), the electrical field operators take the form
\begin{eqnarray}\label{eq:targetfieldop}
  \hat{{\bf E}}_{s,i} &=& \frac{1}{2}[\hat{\bf E}^{(+)}_{s,i}({\bf r},t) +
  \hat{\bf E}^{(-)}_{s,i}({\bf r},t)] \nonumber \\ 
  &=& {i\over 2}\sum_{k_{s,i}} \sqrt{\frac{2 \hbar
        \omega_{s,i}}{n^2_{s,i} \epsilon_0}}\mbox{ }
    \frac{\alpha_{s,i}}{\sqrt{L}} \mbox{ } {\bf e}_{s,i} g_{s,i}({\bf r})\mbox{ } e^{ - i\omega_{s,i}t } \mbox{ } \hat{a}_{k_{s,i}} \nonumber \\ && +\, h.c.  
\end{eqnarray}
Here, ${\bf e}_{s,i}$ indicate the polarization vectors, and $n_{s,i}$
and $\omega_{s,i}$ the corresponding refractive indices and angular
frequencies of the target modes. 
The modes are indexed by the scalar moduli $k_{s,i}$, and the corresponding
full wave vectors in the pump coordinates are given by
\begin{equation}
  {\bf k}_{s,i}=k_{s,i}(\mp \sin\theta_{s,i} {\bf e}_y +
  \cos\theta_{s,i}{\bf e}_z)\,.
\end{equation} 

The longitudinal wave vector components serve as a
complete discrete mode index $k_{s,i}=2\pi m_{s,i}/L$ with integer numbers
$m_{s,i}$. The coefficients before the raising and lowering operators in
eq.~(\ref{eq:targetfieldop}) are chosen such that the free field Hamiltonian
$\hat{H}_0$ for the target modes takes the usual form
\begin{equation}
\hat{H}_{0}=\sum_{k_{s,i}}\hbar\omega_{s,i}\left(
  \hat{a}^{\dagger}_{k_{s,i}}\hat{a}_{k_{s,i}}+{1\over2}\right).
\end{equation}

\subsection{Interaction Hamiltonian}
The SPDC process is enabled by a nonlinear optical material whose presence is
described by the Hamiltonian $\hat{H}_I$, written in the
interaction picture with time dependence of the raising and lowering
operators \cite{shentxt}:
\begin{eqnarray}\label{eq:interactionhamil}
\hat{H_I} &=& -\frac{2 \epsilon_0 \chi^{(2)}}{8} \int
\limits^{\infty}_{-\infty}\!\!\!dxdy\! \int\limits_{-l/2}^{l/2}\!\!\!\!dz\, {\bf E}_p^{(+)} \hat{\bf E}_s^{(-)} \hat{\bf E}_i^{(-)} + h.c.  \nonumber \\ 
&=& {d}\!\!
\int\limits^{\infty}_{-\infty}\!\!\!dxdy\!\!\int\limits^{l/2}_{-l/2}\!\!\!\!dz 
 \sum_{k_s,k_i} \mbox{ } \frac{\hbar \sqrt{\omega_i \omega_s}}{n_s n_i }
   \frac{\alpha_s \alpha_i E^0_p}{L}\nonumber \times \\ && \times \,
  e^{-i\Delta \omega t} g_p({\bf r}) g^{*}_s({\bf r}) g^{*}_i({\bf r}) \hat{a}_{k_s}^\dagger(t)\hat{a}_{k_i}^\dagger(t) +\, h.c.
\end{eqnarray} 
We have assumed a crystal of infinite transverse $(x,y)$ extent, which is justified when the beam diameters are much smaller than the crystal dimensions. 
We introduce a frequency mismatch $\Delta\omega=\omega_p-\omega_s-\omega_i$.
The effective non-linearity $d$ captures the contraction
of the nonlinear susceptibility tensor with the corresponding polarization
vectors ($2 d={\bf e}_p \chi^{(2)}:{\bf e}_{s}{\bf e}_i$) \cite{shentxt}.
With this notation, the type of phase matching condition (type I or II) is
reflected in an appropriate effective non-linearity $d$.

Most of the scaling aspects of the parametric down conversion process
connected with the geometry of the interaction are determined by the overlap
integral $\Phi (\Delta k)$ of the three mode functions $g_{p,s,i}({\bf r})$ in
the crystal:
\begin{eqnarray}\label{eq:overlap1}
\Phi(\Delta {\bf k}) &=& \int\!\!dz\!\int\!\!dy\,dx\, g_p({\bf r}) g^{*}_s({\bf r}) g^{*}_i({\bf r})\nonumber \\
&=& \int\!\!dz\!\int\!\!dy\,dx\,e^{i\Delta {\bf k}\cdot{\bf r}} U_p({\bf r})U_s({\bf r}) U_i({\bf r}). 
\end{eqnarray}
In this expression, $\Delta {\bf k} = {\bf k}_p - {\bf k}_s - {\bf k}_i$ describes the wave vector mismatch.  
Since pump and target modes are defined in the $y$-$z$ plane, there are no wave vector components in the $x$-direction and hence $\Delta k_x =0$.
Carrying out the integration in the 
transverse directions ($x,y$) we arrive at
\begin{equation}\label{eq:overlap2}
\Phi(\Delta {\bf k}) = {\pi\over\sqrt{A\cdot C}}\, 
    e^{-{\Delta k_y^2\over4C}} 
\!\int\!\!dz\,e^{-H z^2+ i z K},
\end{equation}
with the abbreviations
\begin{eqnarray}\label{eq:abbrevs1}
A&=&{1\over W_p^2}+{1\over W_s^2}+{1\over W_i^2}\\
C&=&{1\over W_p^2}+{\cos^2\theta_s\over W_s^2}+{\cos^2\theta_i\over W_i^2}\\
D&=&{\sin2\theta_s\over W^2_s}-{\sin2\theta_i\over W^2_i}\\
F&=&{\sin^2\theta_s\over W_s^2}+{\sin^2\theta_i\over W_i^2}\\
H&=&F-{D^2\over4C}\\
K&=&\Delta k_y{D\over2C}+\Delta k_z
\end{eqnarray}

The exponential term before the residual integral in eq.~(\ref{eq:overlap2}) represents the approximate transverse wave vector mismatch.
This term can be ignored only if one of the beams is infinitely large ($W_{p,s,i}\rightarrow\infty$), or if there is perfect transverse phase matching.

The residual integral along $z$ in (\ref{eq:overlap2}) can be
re-written in a form that allows for a physical interpretation. 
We introduce $\Phi_z$ where
\begin{eqnarray}\label{eq:overlap3}
\Phi_z&:=&\int\limits_{-l/2}^{l/2}\!\!dz\,e^{-Hz^2+i z K}\\
&=&l\cdot\int\limits_0^1\!\!du\,e^{-\Xi^2u^2}\cos (\Delta \varphi u).
\end{eqnarray}
The phase mismatch is now defined as $\Delta \varphi := K l/2$. 
The argument $\Xi :=\sqrt{H} l/2$ in the exponential can be viewed as a
``walk-off" parameter due to non-collinear mode propagation.
This parameter is useful for identifying a thin crystal  and a thick crystal
regime \cite{monken98}. 
In our model, these regimes refer to the physical boundary conditions imposed
on the interaction volume by the geometry of the pump and target modes. 

In the thick crystal regime with a large walk-off parameter ($\Xi > 1$), the overlap integral $\Phi_z$ depends mostly on the characteristic beam parameters $W_{p,s,i}$ and not much on the physical boundaries of the non-linear material.  
For $\Xi \rightarrow \infty$ the length of the crystal ceases to play a role altogether: 
\begin{eqnarray}
\Phi_z &\approx& l {\sqrt{\pi}\over 2 \Xi}\mbox{ Erf}(\Xi) \nonumber \\
&=& \sqrt{{\pi \over H}} \mbox{ Erf}(\Xi) 
\end{eqnarray}

The thin crystal regime refers to a small walk-off parameter, $\Xi\ll 1$, so that the characteristic beam parameters have almost no influence on $\Phi_z$.
In particular, this applies for collinear arrangement of all modes ($\theta_i=\theta_s=0$), where $\Xi=0$. 
In this case, $K=\Delta k_z$, and \begin{equation} \Phi_z=l\,\mbox{sinc}(\Delta \varphi) \end{equation}

\begin{figure} \centerline{\scalebox{0.7}{\includegraphics{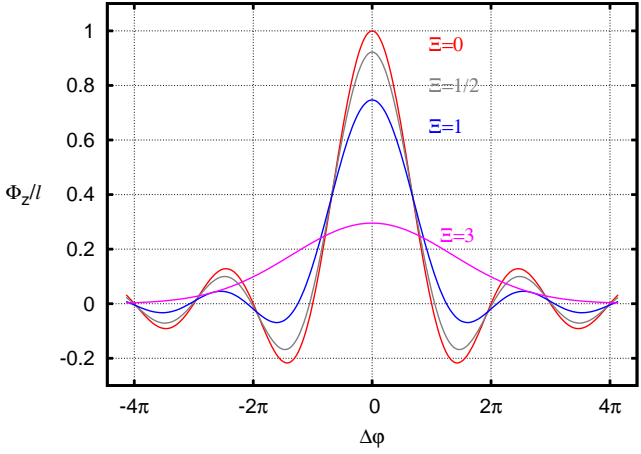}}}
  \caption{\label{fig:zoverlap} (Color online)
    Longitudinal overlap function $\Phi_z/l$ as a
    function of the total phase mismatch $\Delta\varphi=K l/2$ over the
    crystal for various walk-off parameters $\Xi$. For $\Xi=0$, the typical
    sinc shaped spectral distribution is revealed, whereas for large walk-off
    parameters $\Xi>1$ the phase matching condition is determined by the
    overlap region formed by pump- and target modes, and develops into a
    Gaussian distribution.
  }
\end{figure} 

This reveals the well-known influence of the longitudinal phase mismatch on the downconversion spectral properties \cite{theories}.
Figure~\ref{fig:zoverlap} shows the overlap contribution $\Phi_z/l$ as a function of the phase mismatch $\Delta\varphi$ for various walk-off parameters $\Xi$. 
We note that as $\Xi$ becomes large, the spectrum becomes Gaussian-like.  
If $\Xi$ is identified as the degree of (non)collinearity, it suggests that the SPDC spectral profile far from collinear emission can be very broad.
In other words, for the same beam parameters $W_{s,i,p}$, the bandwidth of
collinear emission will be narrower than for non-collinear emission.

\subsection{Spectral Emission Rate }
To obtain absolute emission rates, we make use of Fermi's Golden Rule
as an expression for the transition rate $R(k_s)$ between the initial vacuum
state $\ket{i}=\ket{0_{k_s},0_{k_i}}$, and a final state
$\ket{f}=\hat{a}_{k_s}^\dagger\hat{a}_{k_i}^\dagger\ket{0_{k_s},0_{k_i}}$
with the mode pair $k_s, k_i$ populated with one photon each. 
Fermi's rule applies for asymptotic scattering rates, so the relation between
$k_s$ and $k_i$ is fixed by energy conservation:
\begin{equation}\label{eq:econservation}
\Delta\omega=\omega_p-k_s{c\over n_s}-k_i{c\over n_i}=0
\end{equation}

We first evaluate the transition rate $R(k_s)$ to a {\em fixed signal target
  mode } $k_s$.  The density of states $\rho$ per unit of energy
$\Delta E=\hbar\Delta\omega$ is extracted out of a quasi-continuum of states
for the mode $k_i$:
\begin{equation}
  \rho(\Delta E)=
  {\Delta m\over\Delta k_i}{\partial k_i\over\partial (\hbar\Delta\omega)}
  = {L\over2\pi}{n_i\over\hbar c}\,,
\end{equation}
where $\Delta m/\Delta k_i=L/2\pi$ denotes the number of modes per unit of
wave vector component $k_i$.

With the transition matrix element expressed in terms of the
overlap integral $\Phi(\Delta {\bf k})$,
\begin{equation}
  \bra{f}\hat{H}_I\ket{i}=d\,{\hbar\sqrt{\omega_s\omega_i}\over
      n_sn_i}{\alpha_s\alpha_i\over L} E_p^{0}\Phi(\Delta {\bf k})
    ,,
\end{equation}
the transition rate is then given by
\begin{eqnarray}
  R(k_s)&=&
  {2\pi\over\hbar}\left|\bra{f}\hat{H}_I\ket{i}\right|^2\rho(\Delta E)\\
  &=&\left|{d\mbox{ }\alpha_s\alpha_iE_p^0\Phi(\Delta {\bf
  k}) 
   }\right|^2{\omega_s\omega_i\over n_s^2n_i c L}
\end{eqnarray}

The spectral emission rate per unit of angular frequency $\omega_s$ is
obtained by multiplying $R(k_s)$ with the number of modes  $k_s$ in a unit
interval of $\omega_s$, which is $Ln_s/2\pi c$.  We finally arrive at 

\begin{equation}\label{eq:spectrate}
  {dR(\omega_s)\over d\omega_s} 
= \left[{d\mbox{ }\alpha_s\alpha_i E_p^0 \Phi(\Delta
  {\bf k})\over c} 
   \right]^2{\omega_s\omega_i\over2\pi n_sn_i} \\
\end{equation}

At this point, the earlier introduced quantization length $L$ has vanished as expected. 

\subsection{Total Emission Rate}
We now can determine the total pair generation rate by integrating the
spectral rate density over all frequencies $\omega_s$. Assuming that the
overlap $\Phi(\Delta {\bf k})$ is only non-vanishing over a small range
of frequencies $\omega_s$, the total pair generation rate can be written as  
\begin{equation}\label{eq:rateequation}
  R_T=\left[d\alpha_s\alpha_iE_p^0\over
    c\right]^2{\omega_s\omega_i\over2\pi
  n_sn_i}\int\!\!d\omega_s\left|\Phi(\Delta {\bf k})\right|^2
\end{equation}

The dependency of $\Phi(\Delta {\bf k})$ on $\omega_s$ can be quite involved,
as in the non-collinear case  ($\theta_{i,s}\neq0$) both $\Delta
k_y(\omega_s)$ and $\Delta k_z(\omega_s)$ must be considered.   
However, the alignment criteria for most experimental setups assume perfect longitudinal phase matching to arrive at the collection angles for degenerate downconversion \cite{kwiat95,kurtsiefer01}.  
In these collection directions, the target mode angles $\theta_{s,i}$ are equal.
Furthermore, the typical experiments use identical collection mode diameters
for signal and idler, $W_s=W_i$ \cite{kurtsiefer01}.
Under these two conditions, the phase mismatch $\Delta
\varphi$ depends only on $\Delta k_z$.

We now consider the exponential term for the overlap $\Phi$ in
eq.~(\ref{eq:overlap2}) that contains $\Delta k_y$.
For experiments where light centered on the degenerate wavelengths with a
small bandwidth is collected ($\approx 2$ nm on either side of the center \cite{kurtsiefer01}), we will assume perfect transverse phase matching.  
A complete treatment with non-zero transverse phase mismatch requires a
numerical procedure \cite{boeuf00}. 

With perfect transverse phase matching, we can carry out the integration in
eq.~(\ref{eq:rateequation}) by re-parameterizing the frequencies of the
signal and 
idler about the degenerate SPDC frequency: $\omega_s=\frac{\omega_p}{2} -
\delta_\omega \mbox{ and } \omega_i=\frac{\omega_p}{2}+\delta_\omega$.   
We approximate $\omega_s\omega_i \approx \frac{\omega_p^2}{4}$ by
ignoring terms $O(\delta_w^2)$. 
From energy conservation eq.~(\ref{eq:econservation}) and the phase matching
condition
\begin{equation} \Delta k_z = n_s \omega_s \cos\theta_s + n_i \omega_i\cos\theta_i - n_p\omega_p ,
\end{equation}
 we obtain a dispersion relation between $d\omega_s$ and $d(\Delta k_z)$:
\begin{equation}
d(\Delta k_z) = {(n_i \cos\theta_i- n_s \cos\theta_s)\over c} d\omega_s
\end{equation}

\begin{figure} \centerline{\scalebox{0.35}{\includegraphics{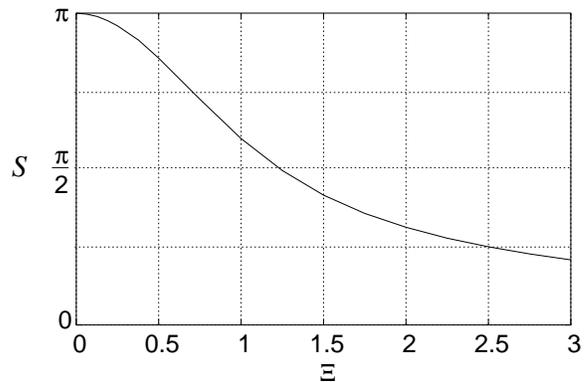}}} \caption{\label{fig:chi_int} Variation of the spectral integral $S$ with walk-off parameter $\Xi$.  Since the absolute emission rate is proportional to $S$, the largest absolute rate is obtained in the thin-crystal limit with  $\Xi=0$.  } \end{figure}

The emission rate can now be integrated over the longitudinal wave vector
mismatch $\Delta k_z$,
\begin{equation}
R_T = 
\frac{d^2 (\alpha_i \alpha_s E_p^0)^2 \omega_p^2 }{ 4 c n_s n_i (2\pi)(n_i\cos\theta_i-n_s \cos\theta_s) } \int |\Phi(\Delta {\bf k})|^2 d(\Delta k_z).
\end{equation}
Effectively, this is the pair emission rate for all allowed wavelengths in the
direction defined by our paraxial beams. 
If we recall that the pump has a Gaussian envelope, and choose all beam
characteristics to be equal ($W_p=W_s=W_i$), the rate $R_T$ finally can be
written as
\begin{eqnarray} \label{eq:finalrate}
  R_T &= &\frac{4d^2  P l\omega_p^2}{3 \pi n_p n_s n_i \epsilon_0 c^2 (\pi
    W_p^2)(1+\cos\theta_i^2 + \cos\theta_s^2)}\times\nonumber\\
  &&\times \frac{1}{(n_i\cos\theta_i-n_s\cos\theta_s)}S,  
\end{eqnarray}
with the spectral integral $S := \int \left |{\Phi_z(\Delta {\bf k_z})\over l} \right |^2 d(\Delta k_z l / 2)$.
The absolute emission rate is proportional to $S$, which has a dependence on
the value of the walk-off parameter $\Xi$ as shown in figure~\ref{fig:chi_int}.  
The spectral integral assumes its largest value $S=\pi$ in the thin crystal
limit. In this limit, closed form expressions for the spectral and total rates are 
\begin{eqnarray} \label{eq:spectralrate} 
  {d\tilde{R}(\omega_s)\over d\omega_s} &=& \frac{2d^2 \omega_p^2 P l^2
    \mbox{sinc}^2(\Delta k_z l/2)}{9\pi n_p n_s n_i \epsilon_0 c^3 (\pi W_p^2)}
  \quad {\rm and}\\
  \label{eq:totalrate}
  \tilde{R}_T &=& \frac{4d^2  P l \omega_p^2}
  {9  n_s n_i n_p\epsilon_0 \pi W_p^2(n_i-n_s) c^2}\,. 
\end{eqnarray}

\subsection{Dependence of Emission Rate on Beam Waists} 
\begin{figure}
  \centerline{\scalebox{0.35}{\includegraphics{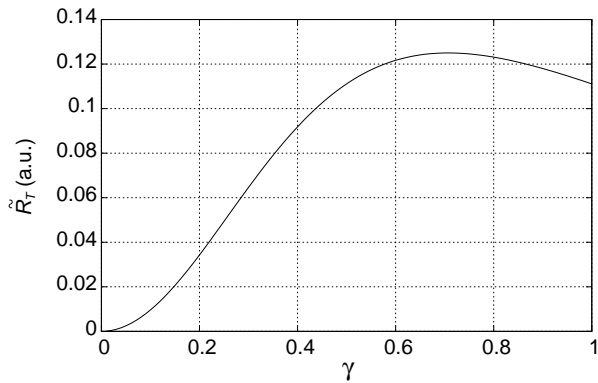}}} 
  \caption{\label{fig:gamma}
    Dependence of the total pair rate $\tilde{R}_T$ on the ratio $\gamma$
    between target and pump waist. The maximum emission rate can be expected
    at $\gamma=1/\sqrt{2}$.
  } 
\end{figure}

Although it is convenient to set all beam waists to be equal, this is not
necessary. 
In fact, it can be shown that this choice does not maximize the total emission
rate for a given optical pump power. Carrying out the more general
derivation to arrive at an expression similar to (\ref{eq:totalrate}), the
dependency on the various beam waists $W_p, W_s$ and $W_i$ (again in the thin
crystal limit) can be written as
\begin{eqnarray} R_T &\propto& \frac{1}{W_p^2 W_s^2 W_i^2 (\frac{1}{W_p^2} +
    \frac{1}{W_s^2} + \frac{1}{W_i^2})^2} \,.
\end{eqnarray}  
To develop an alignment strategy, we may assume that the collection modes are
identical ($W_s=W_i=W$), but we re-express the pump waist as $W_p=\gamma
W$, so we obtain 
\begin{equation} \tilde{R}_T \propto R_T  \propto \frac{1}{W^2
    (\frac{1}{\gamma} + 2\gamma)^2}\,.
\end{equation} 
This relationship is illustrated in figure~\ref{fig:gamma}, and exhibits a
maximum of $\tilde{R}_T$ for $\gamma=\frac{1}{\sqrt{2}}$.  
For $\gamma=1$, the emission rate is about 12\% lower than the maximum value.
This suggests that experimental setups that are designed with equal beam
waists for pump and collection modes may be further optimized, and the simple
argument of maximizing a mode overlap \cite{kurtsiefer01} with matching beam
waists does not hold.

\section{Physical Interpretation and Comparison to Experiments}
While $R_T$ is proportional to crystal length $l$, the spectral rate
$dR(\omega_s)/ d\omega_s$ is proportional to the square of a sinc function.
This is in agreement with results from previous work \cite{theories}.
However, our expression reveals dependencies on other factors, namely emission
geometry and pump spot-size. 
The expression for $R_T$  reveals that the emission rate is higher in a
collinear geometry compared to a non-collinear case. This can be intuitively
understood because the non-collinear case has a smaller interaction volume.  

We find that the absolute rate $R_T$ is proportional to the square of pump
frequency since we re-parameterize the signal and idler about the degenerate
frequency, so downconversion efficiency can be improved with shorter
wavelength pumps. 
Both spectral and total emission rates are inversely proportional to the mode
area of the beams, in contrast to previous studies which showed that the total
SPDC cannot be enhanced by focusing \cite{theories, koch95}. There, however,
SPDC emission was not considered for specific transverse modes.      
The dependence of emission rates on the mode area have been reported in a
previous analysis of SPDC in waveguide structures \cite{fiorentino07}.
This is not surprising because the emission into paraxial beams is essentially
the same problem as SPDC in waveguides, where the target modes are quantized
in one dimension only. 
We note that our equation (\ref{eq:spectralrate}) is similar to the equation
obtained in reference \cite{fiorentino07}.

We should not draw the conclusion, however, that SPDC emission into single transverse modes can be arbitrarily enhanced by tight focusing.
Our model is only valid in cases where the transverse profile of the beams
does not vary significantly over the crystal length. 
For an optimization study of focus size on SPDC emission we refer the reader
to \cite{ljunggren05}.

For explicit comparison of eq.~(\ref{eq:finalrate}) with experimental values, we consider our experimental setup (similar to that used in reference \cite{branciard07}).
In this experiment a pump beam (beam waist, $W_p =82\mu$m) at a wavelength of 351.1 nm is incident on a 2 mm thick $\beta$-Barium Borate (BBO) crystal.  
Two single mode fibers are used to collect degenerate downconverted photons, which is estimated to have an external emission angle of $3.1^{\circ}$.  
The collection modes also have beam waists of $W_{s,i}=82\mu$m.  

Specifically for BBO the effective non-linearity is given by $d=d_{22}\cos^2\theta_c\cos 3\phi_c$.
The angle between pump wave vector and crystal optical axis is $\theta_c=49.7^{\circ}$, while the azimuthal angle is $\phi_c=60^{\circ}$, resulting in an effective nonlinearity of $9 \times 10^{-13}$ m/V ($d_{22}=2.11\times 10^{-12}\mbox{ m}^{-1}\mbox{ V}^{-1}$ according to \cite{klein03}).  
The observed pair rate is approximately 800 pairs $\mbox{mW}^{-1} \mbox{s}^{-1}$ with a pair-to-singles ratio of 0.23.  

The walk-off parameter for this setup is $\Xi=0.933$, indicating that the overlap integral is intermediate between the thin and thick crystal limits.
The largest observable rate according to our model is $2(0.23\times R_T) =
1100 \mbox{ mW}^{-1}\mbox{s}^{-1}$.
The additional factor of 2 is used because in experiments, the geometry is used to collect downconversion emission in two decay paths.

The source of the discrepancy between experiment and our model is hard to identify.  
The assumptions used in the model make it an overestimate, primarily in the re-parameterizing of signal and idler frequencies about the degenerate wavelength.
Experimentally, there are several sources of uncertainty, the main one being the difficulty in establishing pump power very accurately.
For example, the average observed value was arrived by measuring the power
using two different power-meters (a Newport Model 818-UV reported 11.7\,mW
while a Coherent Fieldmaster reported 9\,mW).   
The uncertainty in pump power estimation, however, is not sufficient to make
the observed result compatible with the calculated value. 

According to the model, the conversion efficiency into Gaussian transverse
modes for our experimental setting will be $3\times 10^{-12} \mbox{ mm}^{-1}$
of crystal length.
Other experimentally reported rates in the literature reveal similar downconversion efficiencies  \cite{kurtsiefer01,bellspore,bovino03}.
The total conversion efficiency of SPDC was found to be on the order of $3\times 10^{-8} \mbox{ mm}^{-1} \mbox{ sr}^{-1}$ by Klyshko \cite{klyshkotxt} (for degenerate SPDC with a 500 nm pump wavelength).  Experimentally, our collection angle is $3.3\times
10^{-5}$ sr. If we convert our units to be comparable with Klyshko's result, we obtain an efficiency of $7\times 10^{-8}\, {\rm mm^{-1}\, sr^{-1}}$.

\section{Conclusion}
In conclusion, we have presented expressions that provide absolute
values for the rate and bandwidth of correlated pairs emitted in bulk crystal
SPDC which are in a single Gaussian mode.  These modes may be defined by the
collection profile of single mode fibers, selecting emission in a specific
pair of directions.  The single mode treatment reduces the complexity in the
final expression for the rate equations.

We find that the expression for absolute rates given by the model are slightly larger than experimental observations. 
The model may thus be treated as an idealized case for the total pair emission
rate. 
The small difference between experimentally observed rates and predictions
according to the closed expression for $R_T$, however, suggests that
experimental setups using single mode collection fibers (e.g. \cite{bovino03,
  kurtsiefer01}) operate close to the optimal limit. 

Substantial increase of the emission rates are to be expected from larger non-linearities, since emission rates are proportional to $d^2$.
Small mode diameters are also expected to enhance emission rates, as has been
convincingly reported for SPDC experiments using waveguide structures
\cite{tanzilli01, sanaka01,uren04}, and a similar theoretical analysis
\cite{fiorentino07} for those cases. 
Overall spectral brightness will be improved by combining larger non-linearities with collinear mode confinement in longer structures.
Even then, however, the spectral width is still ultimately determined by the longitudinal wave vector mismatch.
This indicates that very dramatic improvements (by several orders of
magnitude) to the generated pair rate in a narrow bandwidth necessary for
addressing atomic systems is not very likely to be expected from bulk crystal
emission.

\acknowledgments
The authors acknowledge financial support from ASTAR under SERC grant No. 052
101 0043. Special thanks to a referee initiating the discussion on
the beam waist ratios in this paper.


\end{document}